# Possible Contexts of Use for *In Silico* trials methodologies: a consensus- based review

Marco Viceconti, Luca Emili, Payman Afshari, Eulalie Courcelles, Cristina Curreli, Nele Famaey, Liesbet Geris, Marc Horner, Maria Cristina Jori, Alexander Kulesza, Axel Loewe, Michael Neidlin, Markus Reiterer, Cecile F. Rousseau, Giulia Russo, Simon J. Sonntag, Emmanuelle M. Voisin, and Francesco Pappalardo

*Abstract*—The term "*In Silico* Trial" indicates the use of computer modelling and simulation to evaluate the safety and efficacy of a medical product, whether a drug, a medical device, a diagnostic product or an advanced therapy medicinal product. Predictive models are positioned as new methodologies for the development and the regulatory evaluation of medical products. New methodologies are qualified by regulators such as FDA and EMA through formal processes, where a first step is the definition of the Context of Use (CoU), which is a concise description of how the new methodology is intended to be used in the development and regulatory assessment process. As *In Silico* Trials are a disruptively innovative class of new methodologies, it is important to have a list of possible CoUs highlighting potential applications for the development of the relative regulatory science. This review paper presents the result of a consensus process that took place in the InSilicoWorld Community of Practice, an online forum for experts in *in silico* medicine. The experts involved identified 46 descriptions of possible CoUs which were organised into a candidate taxonomy of nine CoU categories. Examples of 31 CoUs were identified in the available literature; the remaining 15 should, for now, be considered speculative.

*Index Terms*— Context of Use, Good Simulation Practice, In Silico Trials.

## I. Introduction

THE term "*In Silico* Trial" indicates the use of computer modelling and simulation to evaluate the safety and efficacy of a medical product, whether a drug, a medical device, or an advanced therapy medicinal product. The evaluation of safety and efficacy has historically been carried out experimentally, using *in vitro* tests, *ex vivo* tests, *in vivo* tests on animals, or *in vivo* tests on humans (clinical trials). But there are various reasons to consider computational models as an integral part of the evidence generation paradigm. *In vitro* and *ex vivo* tests can be time-consuming; animal experimentation is posing growing ethical and translational concerns; clinical trials are very expensive and require considerable time. High costs of innovation and lengthy time-to-market translate into higher selling prices, and under-representation as discrimination (e.g., gender, age, ethnicity, rarity of the condition, wealth). In addition, many of these experimental methods only support rejecting a hypothesis, *e.g.* to show that a product is not safe or

Manuscript submitted on March 27th, 2021. This work was supported in part by the STriTuVaD project (H2020-SC1-2017-CNECT-2/777123), the CompBioMed2 project (H2020-INFRAEDI-2018-1/823712) and InSilicoWorld project (H2020-SC1-DTH-2020-1/101016503).

M. V. is with the Department of Industrial Engineering, Alma Mater Studiorum - University of Bologna, 40136 Bologna IT and the Medical Technology Lab, IRCCS Istituto Ortopedico Rizzoli, 40136 Bologna IT (e-mail: marco.viceconti@unibo.it).

L. E. is with the In Silico Trials Technologies, Via Parini, 9 - 20121 Milano IT (e-mail: luca.emili@insilicotrials.com).

P. A. is with DePuy Synthes Spine, Johnson and Johnson, Raynham, MA, United States (e-mail: pafshari@its.jnj.com).

E. C. is with Novadiscovery SAS, Lyon (FRA) (e-mail: eulalie.courcelles@novadiscovery.com).

C. C. is with the Department of Industrial Engineering, Alma Mater Studiorum - University of Bologna, 40136 Bologna IT and the Medical Technology Lab, IRCCS Istituto Ortopedico Rizzoli, 40136 Bologna IT (e-mail: cristina.curreli@unibo.it).

N. F. is with the Biomechanics Section, Department of Mechanical Engineering, KU Leuven, Leuven, Belgium (e-mail: nele.famaey@kuleuven.be).

L. G. is with the Division of Skeletal Biology & Engineering Research Center, KU Leuven, Belgium; the Biomechanics Section, Department of Mechanical Engineering, KU Leuven, Leuven, Belgium and the Biomechanics Research Unit, University of Liège, Liège, Belgium (e-mail: liesbet.geris@uliege.be).

M. H. is with ANSYS, Inc., Evanston, IL, United States (e-mail: marc.horner@ansys.com).

M. C. J. is with Mediolanum Cardio Research, Milano, Italy (e-mail: jori@mcr-med.com).

A. K. is with Novadiscovery SAS, Lyon (FRA) (e-mail: alexander.kulesza@novadiscovery.com).

A. L. is with the Institute of Biomedical Engineering, Karlsruhe Institute of Technology (DE) (e-mail: axel.loewe@kit.edu).

M. N. is with the Department of Cardiovascular Engineering, Institute of Applied Medical Engineering, RWTH Aachen University (DE) (e-mail: neidlin@ame.rwth-aachen.de).

M. R. is with Medtronic, PLC (USA) (e-mail: markus.w.reiterer@medtronic.com).

C. F. R. is with the Voisin Consulting Life Sciences, Boulogne (Paris) (FRA) (e-mail: rousseau@voisinconsulting.com).

G. R. is with the Department of Drug and Health Sciences, University of Catania (IT) (e-mail: giulia.russo@unict.it).

S. J. R. is with Virtonomy GmbH, Munich (DE) (e-mail: sonntag@virtonomy.io).

E. M. V. is with Voisin Consulting Life Sciences, Boulogne (Paris) (FRA) (e-mail: voisin@voisinconsulting.com).

F. P. is with the Department of Drug and Health Sciences, University of Catania (IT) (e-mail: francesco.pappalardo@unict.it).

not effective, but rarely do they provide information about how such a product can be made safer and more effective.

Evidence of the safety and efficacy of a new medical product is necessary to obtain marketing authorisation from competent authorities (regulatory agencies, notified bodies, etc.), which every company needs before a medical product can be sold in a certain country. All methodologies used to produce evidence of safety or efficacy must be considered appropriate by the competent authorities to whom the evidence is submitted. Such evaluation is usually conducted by the Competent Authorities and is called *certification* or *qualification* (hereinafter we will use the generic term *approval*). Likewise, *In Silico* Trial methodologies must receive the necessary approval before they can be used.

The first step in the regulatory approval of every new methodology is an accurate definition of the Context of Use (CoU)[1]. According to FDA-CDER "The CoU is a concise description of the drug development tool's specified use in drug development". FDA-CDRH defines CoUs as "a statement that fully and clearly describes the way the Medical Device Development Tool is to be used and the medical product development-related purpose of the use". EMA defines the CoU as "Full, clear and concise description of the way a novel methodology is to be used and the medicine development related purpose of the use". In the technical standard V&V 40, the CoU is a statement that defines "the specific role and scope of the computational model used to address the question of interest"[1].

In general, the approval is specific for that CoU, and therefore, a new approval must be granted for each new CoU. *In Silico* Trials, being a new class of methodologies, do not have a clear pathway for regulatory assessment and approval in many cases. To address this, the healthcare community is working towards the definition of Good Simulation Practice (GSP), which are quality standards that will define how to assess and approve an *In Silico* Trial methodology before it can be used to produce regulatory evidence on the safety and efficacy of a new medical product. The first step in the definition of the GSP is the compilation of all possible CoUs for *In Silico* Trials methodologies. However, this list must offer a certain level of generalisation to be useful, hence we need an effective taxonomy for *In Silico* Trials CoUs.

This review paper summarises the results of a consensus process which produced one possible taxonomy for *In Silico* Trials CoUs and a list of 46 possible CoUs, which cover all categories of the proposed taxonomy. Examples of 31 CoUs were identified in the available literature; the remaining 15 should, for now, be considered speculative.

## II. THE CONSENSUS PROCESS

The Avicenna Alliance, a global non-profit organization that brings together healthcare stakeholders with the goal of making *in silico* medicine standard practice in healthcare, established a GSP Task Force in January 2020, to develop a Green Paper on Good Simulation Practice. This paper will include a collection of experts' recommendations that might support and help the future elaboration of GSP standards by standardisation bodies.

The GSP Task Force agreed to develop the Green Paper using a grass-roots consensus process that would be run within a recently established online Community of Practice dedicated to *In Silico* Medicine and Digital Health: the *In Silico World* Community of Practice[2]. The membership is open to any individual who has a professional or educational interest in *In Silico* Medicine and Digital Health. To date, the community of practice provides a discussion forum to 308 global experts, including specialists from academia, industry, competent authorities, software houses, etc.

After some preliminary activities, the working group started the CoU collection early April 2020. A first draft list of CoUs was considered too broad and unfocused. Thus, it was agreed to narrow the attention to only those CoUs that have regulatory relevance. On the other hand, given the rapid evolution of this field, it was agreed to also include CoUs that are at the moment only speculative (e.g., for which not a single example can currently be provided, yet).

The first draft of the collection was circulated in late April 2020 and included 33 CoUs; the sixth and final version was circulated to the consensus group in November 2020 and included 46 CoUs. The full list is provided below; where at least one example of that CoU could be identified in the literature, it is included. The CoU without any exemplar reference should be considered speculative for the time being.

## III. THE TAXONOMY FOR CONTEXTS OF USE

The regulatory process aims to provide *Marketing Authorisation* only to those medical products for which the applicant can demonstrate justified claims of safety and efficacy and/or performance. Depending on the type of medical product and its risk class, such claims must be corroborated with evidence of safety and efficacy obtained with a set of controlled experiments conducted *in vitro* or *ex vivo*, *in vivo* on animals, or *in vivo* on humans, with multiple clinical trials involving progressively growing numbers of participants. The controlled experiments in general must be executed in a prescribed order, and each experiment supports the design and provides evidence for the authorisation of the subsequent experiment, or of the final Marketing Authorisation.

*In Silico* Trials aim to reduce, refine, or replace these experiments:

- **Reduce** means to reduce the number of *in vitro* experiments or those involving living subjects (animals or humans), their duration, or the number of experimental subjects (animals or humans) involved in the experiment, or the number of measurements performed during the experiment.
- **Refine** means to revise the study design in order to eliminate or relieve the suffering of the animals involved, or the risks for the humans involved in the experiments; or to shift the experiment to non-animal species, in accordance with animal experimentation ethics. For *in vitro* experiments and animal experiments, refine also means improving the ability of the experiment to predict the results of the human experimentation.

---

[1] A CoU is distinct from the Indication For Use (IFU), which specifies the disease/population for which the medical product is intended.

[2] https://insilico.world/community/

- **Replace** means to replace entirely the experiment, whether *in vitro*, *ex vivo* or *in vivo* in animals or humans, with computational models and simulations.

Thus, the Context of Use for *In Silico* Trials can be taxonomized using the three-by-three matrix depicted in table 1.

TABLE I
TAXONOMY OF THE POSSIBLE CONTEXTS OF USE OF *IN SILICO* TRIAL METHODOLOGIES.

|  | Reduce | Refine | Replace |
|---|---|---|---|
| *Preclinical In Vitro/Ex Vivo Experiments* | Reduce the number or duration of in vitro/ex vivo experiments | Improve the predictive accuracy of safeness and/or effectiveness provided by the in vitro or ex vivo experiment | Replace entirely a portion or all the required in vitro or ex vivo experiments |
| *Preclinical Animal Experiments* | Reduce the number of animals involved in the experiment, or its duration | Alleviate the suffering of the animals involved, or improve the predictive accuracy of the safeness and/or effectiveness provided by the animal experiment | Replace animal experiments in the prediction of the expected safety and/or efficacy for a new treatment during the clinical experimentation |
| *Clinical Human Experiments* | Reduce the number of humans involved in the experiment, or its duration | Reduce the risks for the humans involved, or improve the predictive accuracy of the safeness and/or effectiveness provided by the human trials | Replace human experiments in the prediction of the expected safety and/or efficacy for a new treatment during real-world, post-marketing use |

## IV. IN VITRO / EX VIVO EXPERIMENTS

### A. Reduce In Vitro or Ex vivo experiments

1. *In silico* models can provide a framework for mapping measurements from one experimental setup to another. This can be useful when multiple measurements are made in earlier experiments. The model can then be used to guide and prioritise future experiments. **Example:** Ion channel inhibition in an expression system is integrated into stem cells or animal cells [2].
2. For *in vitro or ex vivo* experiments aimed to support safety assessment, *in silico* models can identify the worst-case scenario relevant to the intended clinical use in the target population to be tested experimentally. **Example:** A model predicts increased blood damage of rotary blood pumps when used in low-flow operating conditions [3].
3. For *in vitro* experiments where the effect of a large number of parameters needs to be explored, such exploration can be carried out *in silico*, reducing the number of experiments required to validate the model. **Example:** An *in vitro/in silico* framework for off-target receptor toxicity of chemicals [4].

### B. Refine In Vitro or Ex vivo experiments

4. *In silico* models can identify/prioritise sources of variability to be minimized in an experimental design. **Example:** A model predicts that cell-to-cell adhesive heterogeneity in a tumour population influences Glioblastoma invasion and needs to be considered in i*n vitro* experiments [5].
5. *In silico* models can identify biomarkers that best measure a value of interest. **Example:** Machine learning reveals serum sphingolipids as cholesterol-independent biomarkers of coronary artery disease [6].
6. A specimen-specific *in silico* model prediction can be used as a surrogate biomarker in an experimental study, when that value of interest is difficult, risky, too expensive or impossible to be observed experimentally. **Example:** A computational tool that can simulate the behaviour of a population of cells cultured in a 3D scaffold [7].
7. *In silico* models can be used to support the construct validity of a specific biomarker, by demonstrating convergent validity, separation between known groups, and/or discriminant validity. **Example:** A model predicts that protein kinase identified as mediator of resistance in thyroid cancer stem cells after BRAF inhibitor treatment [8].
8. *In silico* models can provide evidence that supports the selection of drug dosage, device configuration, delivery mode, etc. to be used in the clinical trials. **Example:** *in silico* tumor model to predict responses of Glioblastoma cells to targeted drugs and use this model to stratify patients for clinical trials [9].
9. *In silico* models can confirm or refute the declared mechanism of action for a new medical product. **Example:** Identifying drug effects as alterations of cell signalling pathways through *in silico* modelling [10].

### C. Replace In Vitro or Ex vivo experiments

10. *In silico* models can replace *in vitro or ex vivo* experiments when it is demonstrated that they provide an equivalent or more realistic representation of the clinical conditions of use than any currently available *in vitro or ex vivo* experiment. **Example**: a new treatment for Cryptosporidiosis, after some preliminary results *in vitro*, was optimised for potency *in silico* [11].

## V. ANIMAL EXPERIMENTS

### A. Reduce animal experiments

11. Specific *in silico* models of a wide range of different animal models and breeds can better define the relevant animal type and size prior to *in vivo* testing. **Example**: an *in silico* methodology was used to optimise the size and the positioning of the Realheart total artificial heart in bovines [12].

12. Where animal-specific *in silico* models can quantify certain biomarkers as accurately as in the animal experiment, *in silico*-augmented animal studies can reduce the number of animals required to achieve statistical significance. **Example:** Simulation of thrombogenicity in ventricular assist devices correlates with *in vitro* and *in vivo* animal studies [13].
13. *In silico* models can extrapolate the long-term results of a longitudinal animal experiment, reducing the duration of the study. This also reduces the numerosity by lowering the animals lost at follow-up. **Example**: a statistical model is used to predict the effect of attrition (loss of animals) in longitudinal studies [14].
14. Use *in silico* models to interpolate intermediate results when the investigation requires a time-course study. **Example:** use a model to interpolate between in vivo microCT scans in longitudinal studies on mice, to reduce the x-ray exposure, which could affect the animal experiment [15].
15. In animal experiments where the effect of a large number of parameters needs to be explored, such exploration can be performed *in silico*, reducing the number of experiments required for the nonclinical validation of the medical product. **Example:** *in silico* tools for evaluating rat oral acute toxicity [16].

### B. Refine animal experiments

16. In preparation for a clinical trial, the use of *in silico* models can improve the translation of results from the animal model to humans: **Example**: an *in silico* model validated with animal experiments is used to optimise the delivery protocol and maximise efficacy in the development cancer preventive vaccines [17].
17. *In silico* models can be used for detailed retrospective evaluation of *in vivo* experiments based on the data generated in case an intervention has failed.
18. *In silico* models can provide virtual control arms, replacing steps that involve animal suffering, such as sham operations.
19. In animal models where the disease is induced surgically, use the animal experiment to observe the response to treatment in wild-type animals, and then use an *in silico* model to simulate the change in response when the disease is present.

### C. Replace animal experiments

20. *In silico* models can replace animal experiments when they provide an equivalent or more realistic representation of the clinical conditions of use than any currently available animal experiment. **Example:** Human *in silico* drug trials demonstrate higher accuracy than animal models in predicting clinical pro-arrhythmic cardiotoxicity [18].
21. Combine *in silico* models with cell-based biological set-ups to obtain relevant testing conditions and map cellular system results to appropriate cell type, tissue type or organ type observations, as a means to replace entirely animal experimentation. **Example:** a recent review on these approaches is available [19].
22. *In silico* models can replace *in vivo* experiments for interventional training through realistic and immersive real-time simulation. **Example**: while we could not find direct comparisons, there are studies that use *in silico* training and other studies that use in vivo training on animals [20], [21].

## VI. HUMAN TRIALS

### A. Reduce human experiments

23. *In silico* models can determine, justify, and/or confirm eligibility or exclusion criteria for proper patient or treatment selection. Consolidate heterogeneous patient populations by providing *in silico* evidence of the irrelevance of potential confounding factors. **Example**: use a digital twin model to decide which abdominal aortic aneurysm patients need treatment. Such technologies can also be use to refine the inclusion criteria of a clinical trial [22].
24. Use an *in silico* model to enrol the right patients for a clinical trial and evaluate the suitability of each patient in detail prior to treatment.
25. Patient-specific *in silico* model (a model informed with some quantifications made on an individual patient) that provides a quantification of a biomarker more accurately and precisely than its experimental quantification, reducing the number of patients required to achieve statistical significance. **Example**: in a clinical trial of a new thrombectomy device, the primary end point, the volume of saved brain tissue was determined using a using a machine learning model [23].
26. Extrapolate the long-term results of a longitudinal clinical trial, reducing the duration of the study. This also reduces the numerosity by lowering the patients lost at follow-up.
27. Patient-specific *in silico* model that provides a quantification of a biomarker that is used as primary endpoint in the clinical trial; the trial is powered to ensure significance on the discrimination for this biomarker. **Example:** An MRI-based patient-specific cartilage degeneration algorithm was able to predict osteoarthritis (OA) progression closely to that observed radiographically [24].
28. Use a translational *in silico* model informed by preclinical experiments to inform clinical development decisions. **Example**: model based on cellular signalling interactions that predict the influence of a cytokine on the survival, duplication and differentiation of the CD133+ HSC/HPC subset from human umbilical cord blood [24].
29. Use Bayesian adaptive clinical trial designs, where the prior is provided by *In Silico* Trials on virtual patients (*In Silico*-Augmented Clinical Trials) to reduce the number of enrolled patients. **Example:** A novel method based on the power prior for augmenting a clinical trial using virtual patient data is illustrated by a case study of cardiac lead fracture [25].
30. Use *In Silico*-Augmented Clinical Trials to achieve statistical significance when there is sparsity in the clinical data, to include as virtual patients sub-groups that are difficult to recruit, or when the recruitment in general is difficult, as is frequently the case for rare diseases.
31. Where such effects are well described and can be modelled, use *In Silico* Trials to account for response biases due to biological sex, gender, ethnicity, lifestyle, etc., thereby reducing the size of the clinical trial required to account for all these subgroups. **Example:** Design of electrocardiography criteria in a retrospective cohort (n = 76) and then analysed in a validation cohort (n = 53) [26].

*B. Refine human experiments*

32. *In silico* models that provide virtual control arms, to avoid enrolling healthy volunteers, replace placebo arms when the ethics of non-treatment is questionable, or when the control arm involves unnecessary risks for the participants. **Example:** A statistical model for a virtual control arm for chemotherapy [27].
33. Enrich existing clinical trial results by simulating unexplored scenarios (*e.g.*, additional regimen) and/or predicting an additional outcome. **Example**: find *in silico* that a specific risk (e.g. restenosis) can be reversed spontaneously over time, in tissue-engineered vascular graft for paediatric use [28].
34. Shorten the learning curve for a specific intervention through *in silico* training. **Example**: Use of in silico training for craniofacial, hand, microvascular, and aesthetic surgery [29].
35. *In Silico* Trials conducted on virtual populations to define clinical trial inclusion and exclusion criteria. **Example:** *In silico* trials for acute ischemic stroke through virtual clot and patient populations [30].
36. *In silico* models can identify the root cause of failures that may occur under broad clinical use. **Example**: A model predicts the potentially differential effects of atrial dilation vs. hypertrophy on the ECG P-wave [31].
37. The inclusion of virtual patients to enrich existing clinical trials with *In Silico* Trials for underserved or underrepresented populations (*e.g.*, paediatric, rare disease) and to accelerate clinical trials when enrolment is challenging.
38. *In Silico* Trials that support Post-Marketing Surveillance by using *in silico* evidence to provide accelerated marketing authorisation. This approach is constrained by a tight post-marketing surveillance, where outcomes and adverse effects must not deviate from the statistical distribution predicted in silico.

*C. Replace human experiments*

39. *In Silico* Trials used as surrogate human experiments when these are impossible to perform.
40. *In Silico* Trials to replace phase II clinical trials to evaluate dose-response in patients. **Example**: *in silico* evaluation of ivabradine efficacy in patients with angina pectoris [32].
41. *In Silico* Trials for phase jumping. For example, the use of phase II clinical trial results to predict the results of a phase III clinical trial, as evidence to support the request for conditional marketing approval (post-marketing phase III trial).
42. *In Silico* Trials of a product that already received marketing authorisation, to validate relabelling, repurposing, retargeting, minor design modifications, etc. This includes also relabelling for paediatric use.
43. Safety *In Silico* Trials to assess possible interactions between multiple treatments.
44. Efficacy *In Silico* Trials of multi-product treatments.
45. *In Silico*-only trials for lower risk class medical devices.
46. *In Silico* Trials for *in vitro* diagnostic medical devices.

## VII. CONCLUSIONS

This review paper summarised the results of a consensus process which produced one possible taxonomy for In Silico Trials CoUs and a list of 46 possible CoUs, which cover all categories of the proposed taxonomy.

The authors believe that the consensus on possible Contexts of Use as the one proposed here is a first essential step in the development of Good Simulation Practice for In Silico Trials. Such a list allows for discussions on the regulatory science behind the use of In Silico Trials that are grounded into a concrete spectrum of use cases.

Although we have collected 46 different use cases, and many more could be defined at a lower level, the same principle of reduce, refine, replace can be applied to the generation of regulatory evidence that is typically done using *in vitro*, animal, or human experimentation.